\documentclass[11pt,a4paper]{article}
\usepackage{jcappub}

\usepackage{comment}
\usepackage{slashed}
\usepackage{color}

\definecolor{red}{rgb}{1,0,0}
\definecolor{darkred}{rgb}{0.6,0,0}
\definecolor{darkgreen}{rgb}{0.992447,0.623778,0.034597}
\definecolor{ppink}{rgb}{1,0.4,0.4}
\definecolor{bblue}{rgb}{0.284602,0.317763,0.963947}

\newcommand{\Mpl}{M_{\rm Pl}}

\newcommand{\Min}{{\rm Min}}
\newcommand{\Max}{{\rm Max}}

\def\Mpl{M_{\rm Pl}}
\def \hc{\rm{h.c.}}

\newcommand{\ltsim}{\protect\raisebox{-0.5ex}{$\:\stackrel{\textstyle <}{\sim}\:$}}
\newcommand{\gtsim}{\protect\raisebox{-0.5ex}{$\:\stackrel{\textstyle >}{\sim}\:$}}

\newcommand{\footnoteref}[1]{\protected@xdef\@thefnmark{\ref{#1}}\@footnotemark}

\title{Primordial Black Holes from Affleck-Dine Mechanism}

\author[a,b]{Fuminori Hasegawa}
\author[a,b]{Masahiro Kawasaki}
\affiliation[a]{ICRR, University of Tokyo, Kashiwa, 277-8582, Japan}
\affiliation[b]{Kavli IPMU (WPI), UTIAS, University of Tokyo, Kashiwa, 277-8583, Japan}
\emailAdd{fuminori@icrr.u-tokyo.ac.jp}
\emailAdd{kawasaki@icrr.u-tokyo.ac.jp}

\abstract{
The recent observations of the gravitational waves (GWs) by LIGO-Virgo collaboration infer the increasing possibility of the primordial black holes~(PBHs).
Recently it was pointed out that sufficient PBHs are produced by the Affleck-Dine mechanism where inhomogeneous baryogenesis takes place due to change of the Hubble induced mass during and after inflation and forms high baryon bubbles (HBBs).
The produced HBBs have large density contrasts through the QCD phase transition or stable Q-ball formation, which leads to formation of the LIGO PBHs.
Furthermore, in this model stringent constraints from CMB $\mu$-distortion and pulsar timing array~(PTA) experiments are completely absent.
In this paper, we study the model in full details based on gravity and gauge mediated supersymmetry breaking scenarios and show that the model can explain the current GWs events evading observational constraints.
}


\begin{document}

\begin{flushright}
    IPMU18-0115
\end{flushright}
\maketitle

\section{Introduction}
The recent observations of the gravitational waves from the merger of the binary black holes~(GW150914~\cite{TheLIGOScientificCollaboration2016b}, GW151226~\cite{TheLIGOScientificCollaboration2016a}, GW170104~\cite{TheLIGOScientificCollaboration2017a}, GW170814~\cite{Abbott2017}) have reveled the existence of the heavier BHs with a mass $\sim 30M_\odot$.
It is in dispute how such heavy BH binaries are formed by the stellar evolution, and many researchers are exploring their origin.
A primordial black hole (PBH) is one of the candidates which account for these heavier BHs~\cite{Bird2016,Clesse2016,Kashlinsky:2016sdv,Carr2016,Eroshenko2016,Sasaki2016}.
PBHs are formed by the gravitational collapse of the over-dense regions in the early Universe.
Therefore, on the contrary to the stellar ones, PBHs can have a very wide range of masses relating to the scale of the over-dense regions.
To generate such over-density in the early universe, inflation is well-motivated and studied extensively~\cite{Yokoyama:1995ex,GarciaBellido:1996qt,Kawasaki:1997ju,Kawasaki2016,Inomata2016,Inomata2017aa,Inomata2017}.
Since the density perturbation generated by the conventional inflation is predominantly scale invariant and too small to collapse to the PBHs, much effort has been made to amplify the curvature perturbations only at the small scale corresponds to $\sim 30M_\odot$.

Meanwhile, such amplified small scale density perturbations which can be the seeds of the PBHs are severely constrained by the cosmological observations.
First, they cause a distortion of the Cosmic Microwave Background~(CMB) spectrum because these perturbations dissipate into the background thermal plasma.
In fact, the observation of the $\mu$-distortion excludes the density perturbations which correspond to the PBHs with mass $4\times10^2M_\odot\lesssim M_{\rm PBH}\lesssim4\times10^{13}M_\odot$~\cite{Kohri2014}.
Moreover, large curvature perturbations could source the tensor perturbations by the second-order effect~\cite{Saito2008,Saito2009}.
It is known that such secondary GWs can be significantly larger than those of the first-order and severely constrained by observations of pulsar timing.
According to the latest results of pulsar timing array~(PTA) experiments~\cite{Arzoumanian2015,Lentati2015,Shannon2015}, the inflationary PBHs with mass $0.1M_\odot\ltsim M_{\rm PBH}\ltsim10M_\odot$ are already excluded.
Consequently, PBHs can explain the massive BHs only in limited mass range as long as curvature perturbations are used as their seeds.
Fortunately, there still exist some successful models which consistently explain the LIGO events evading those constraints by inflationary PBHs~\cite{Inomata2016,Inomata2017}.


As an alternative mechanism to create the PBHs, the one which utilizes the Affleck-Dine~(AD) baryogenesis~\cite{Affleck1985,Dine1996} was studied in the ref.~\cite{Dolgov1993,Dolgov2008,Blinnikov2016}.
In this mechanism, baryon asymmetry is inhomogeneously generated and the localized high-baryon regions called ``high-baryon bubbles (HBBs)" are created.
Since the HBBs become over-dense in the subsequent cosmological evolution, they could gravitationally collapse into the PBHs.
In their setup, however, the ad-hoc and non-SUSY interactions between the inflaton and the AD field are required although the AD baryogenesis is most naturally realized in the framework of SUSY.

Recently, we found that this inhomogeneous AD-baryogenesis can be naturally embedded into the SUSY setup where the dynamics is described by the MSSM flat-directions \cite{Hasegawa:2017jtk}.
The point is that the thermal potential for the AD field, which is originated from thermal plasma produced by the inflaton decay, can trigger the multi-vacua structure of the scalar potential and hence the baryon asymmetry is inhomogeneously produced.
Since this mechanism does not require large Gaussian curvature perturbations, the stringent constraints from the  $\mu$-distortion and PTA experiments are completely absent. Furthermore, it is revealed that the PBHs which explain the LIGO events and the dark matter can be cogenerated if Q-balls~\cite{Coleman1985,Enqvist1998,Enqvist1999,Kasuya2000g,Kasuya2000,Kasuya2001} are formed after the AD baryogenesis.

In this paper, we discuss the PBH formation from the AD baryogenesis more detail and its possible extension. We consider both gravity- and gauge- mediated SUSY breaking scenarios where the properties of the Q-balls formed after the AD baryogenesis are different.
In the gravity-mediated SUSY breaking scenario, since Q-balls are unstable against decay into the nucleons, the HBBs become dense due to the QCD phase transition which transfers the relativistic energy of the massless quarks into the non-relativistic energy of the baryons.
On the other hand, Q-balls in the gauge-mediated SUSY breaking scenario are stable.
Therefore, their non-relativistic energy density eventually dominates the HBBs making the HBBs more dense than the outside the HBBs. Since the HBBs should have high baryon density such as $n_b/s\sim1$ in both cases, we also examine whether AD baryogenesis can really realize these situation for natural cosmological parameters.
The HBB scenario in a single inflation model generally predicts a significant number of the residual HBBs which could cause the cosmological problems.
Therefore, we consider HBB formation in a double inflation model which enables us to suppress the excessive residual HBBs.

This paper is organized as follows.
In the Sec.~\ref{sec:HBB}, we demonstrate the outline of the scenario and discuss how the HBBs are produced from the AD baryogenesis.
In the Sec.~\ref{sec:HBB_ditribution}, the distribution of the HBBs over their scales is evaluated in detail. We explain how the HBBs evolve and gravitationally collapse into the PBHs in the Sec.~\ref{sec:gravitatinal_collapse}.
In the Sec.~\ref{sec:LIGO_event}, we evaluate the abundance of the PBHs comparing with the current observational bounds and discuss its consistency with the current observation of the GWs.
The extension to the double inflation scenario is performed in the Sec.~\ref{sec:double_inflation}.
The Sec.~\ref{sec:conclusion} is devoted to conclusions and discussions.



\section{HBBs from Affleck-Dine field}
\label{sec:HBB}

In this section, we consider the generation of the inhomogeneous baryon asymmetry, that is, HBBs.
In the previous work~\cite{Hasegawa:2017jtk} it was shown that HBBs are naturally produced in the modified version of the AD baryogenesis.
Before we discuss this mechanism in detail, let us briefly review the conventional frame work of the AD baryogenesis.

\subsection{Conventional Affleck-Dine baryogenesis: a review}

In the MSSM there exist flat directions with non-vanishing $B-L$ charge, called Affleck-Dine fields.
Although the potential of the AD fields is flat in the exact-SUSY limit, it is lifted up by SUSY breaking and non-renormalizable terms with cutoff of the Planck scale $M_\text{Pl}$.
The non-renormalizable term is written as
\begin{align}
   W_{\rm AD}=\lambda\frac{\Phi^n}{n\Mpl},
\end{align}
where $\Phi$ is an Affleck-Dine (super)field, $\lambda$ is a coupling constant and $n(>4)$ is a certain integer determined by specifying a flat direction.
In combination with the SUSY-breaking effect, the scalar potential for the AD field $\phi=\varphi e^{i\theta}$ is given by
\begin{align}
   V(\phi)&= (m_\phi^2-cH^2)|\phi|^2+V_{\rm NR},\\
   V_{\rm NR}&=\left(\lambda a_M\frac{m_{3/2}\phi^n}{n\Mpl^{n-3}}+\hc\right)
      +\lambda^2\frac{|\phi|^{2(n-1)}}{\Mpl^{2(n-3)}},
\end{align}
where $m_\phi$ is the soft SUSY breaking mass and $-cH^2$ is the Hubble induced mass with $H$ being the Hubble parameter.
The first part of $V_\text{NR}$ represents so-called A-term which violates baryon number.
In the conventional scenario, the sign of the Hubble induced mass term, which is determined by the coupling to the inflation sector, is assumed to be negative.
Then, in the high energy regime $H\gtrsim m_\phi$, AD field develops a non-vanishing VEV such that
\begin{align}\label{ADV}
   \phi(t)\simeq
      \left(\sqrt{\frac{c}{\lambda^2(n-1)}}H(t)\Mpl^{n-3}
         \right)^{\frac{1}{n-2}}.
\end{align}
As the Hubble parameter $H(t)$ decreases due to the cosmic expansion, the field value of the AD field $\phi(t)$ also decreases in time.
However, when $H(t)$ becomes smaller than $m_{\phi}$, the effective mass of the AD field flips to positive and AD field starts to oscillate around the origin.
At the same time, the phase direction of the AD field $\theta$ receives a ``kick" from the A-term potential.
Then the AD field $\phi$ starts to rotate in the complex plane, which yields the baryon asymmetry of the universe since the baryon number density is defined by
\begin{align}
   n_b=-2q_b\varphi^2\dot\theta.
\end{align}
After some calculation, we can obtain the accurate expression of the baryon abundance $\eta_b~(=n_b/s)$ by the model parameters as
\begin{align}\label{CAD}
\eta_b&\simeq\epsilon \frac{T_Rm_{3/2}}{H_{\rm osc}^2}\left(\frac{\varphi_{\rm osc}}{\Mpl}\right)^2,\\
\epsilon&=\sqrt{\frac{c}{n-1}}\frac{q_b|a_M|\sin{(n\theta_0+\arg(a_M))}}{3\left(\frac{n-4}{n-2}+1\right)},
\end{align}
where the subscript ``osc" denotes the value evaluated at the time AD field starts to oscillate.

\subsection{Inhomogeneous Affleck-Dine baryogenesis}

The mechanism of the HBB production was first proposed in ref.~\cite{Dolgov1993} and developed in refs.~\cite{Dolgov2008,Blinnikov2016}, where AD baryogenesis is employed  in producing baryon asymmetry inside the HBBs.
The inhomogeneity of the baryon asymmetry is explained by the perturbation of the initial value of the AD field owing to the quantum fluctuations during inflation.
However, their setup is non-SUSY and requires ad-hoc couplings between the inflaton and the AD field. Furthermore, the distribution of the HBBs are hard to be estimated due to the complexity of the setup.
Recently, in ref.~\cite{Hasegawa:2017jtk}, we have succeeded in embedding this mechanism into SUSY, where the dynamics are naturally described by the MSSM flat-directions.
In the model, the distribution of the HBBs is easily evaluated by the model parameters.
Let us see how the HBBs are produced from the MSSM flat directions.

\subsubsection{Model setting}

Although the mechanism is based on the conventional AD baryogenesis discussed above, we put two unconventional assumptions:
\begin{itemize}
\item[(i)] During inflation the AD field has a positive Hubble induced mass, while it has negative one after inflation.
\item[(ii)] After inflation, the temperature $T$ of the thermal bath due to the decay of the inflaton overcomes the Hubble parameter $H$.
\end{itemize}
These assumptions are easily satisfied by appropriately choosing the model parameters.
Under these assumptions, the scalar potential for the AD field is altered and given by
\begin{align}\nonumber
    V(\phi)=
       \begin{cases}
          (m_\phi^2+c_IH^2)|\phi|^2+V_{\rm NR},
          &({\rm during~inflation}) \\
          (m_\phi^2-c_MH^2)|\phi|^2+V_{\rm NR}+V_{\rm T}(\phi),
          & ({\rm after~inflation})
       \end{cases}
\end{align}
where $c_I,~c_M$ are dimensionless positive constants,\footnote{
Such a change of the constant $c$ naturally takes place in the supergravity-based inflation models where a multiple superfields is employed.}
$V_{\rm T}$ is the thermal potential for the AD field induced by the decay product of the inflaton and written as
\begin{align}\label{TP}
 V_{\rm T}(\phi)&=
   \begin{cases}
      c_1T^2|\phi|^2, & f_k|\phi|\ltsim T, \\
      c_2 T^4\ln\left(\frac{|\phi|^2}{T^2}\right),& |\phi|\gtsim T,
   \end{cases}
\end{align}
where $c_1,~c_2$ are $\mathcal{O}(1)$ parameters which are related to the couplings of the AD field to the thermal bath.

\begin{figure}[t]
   \centering
   \includegraphics[width=85mm]{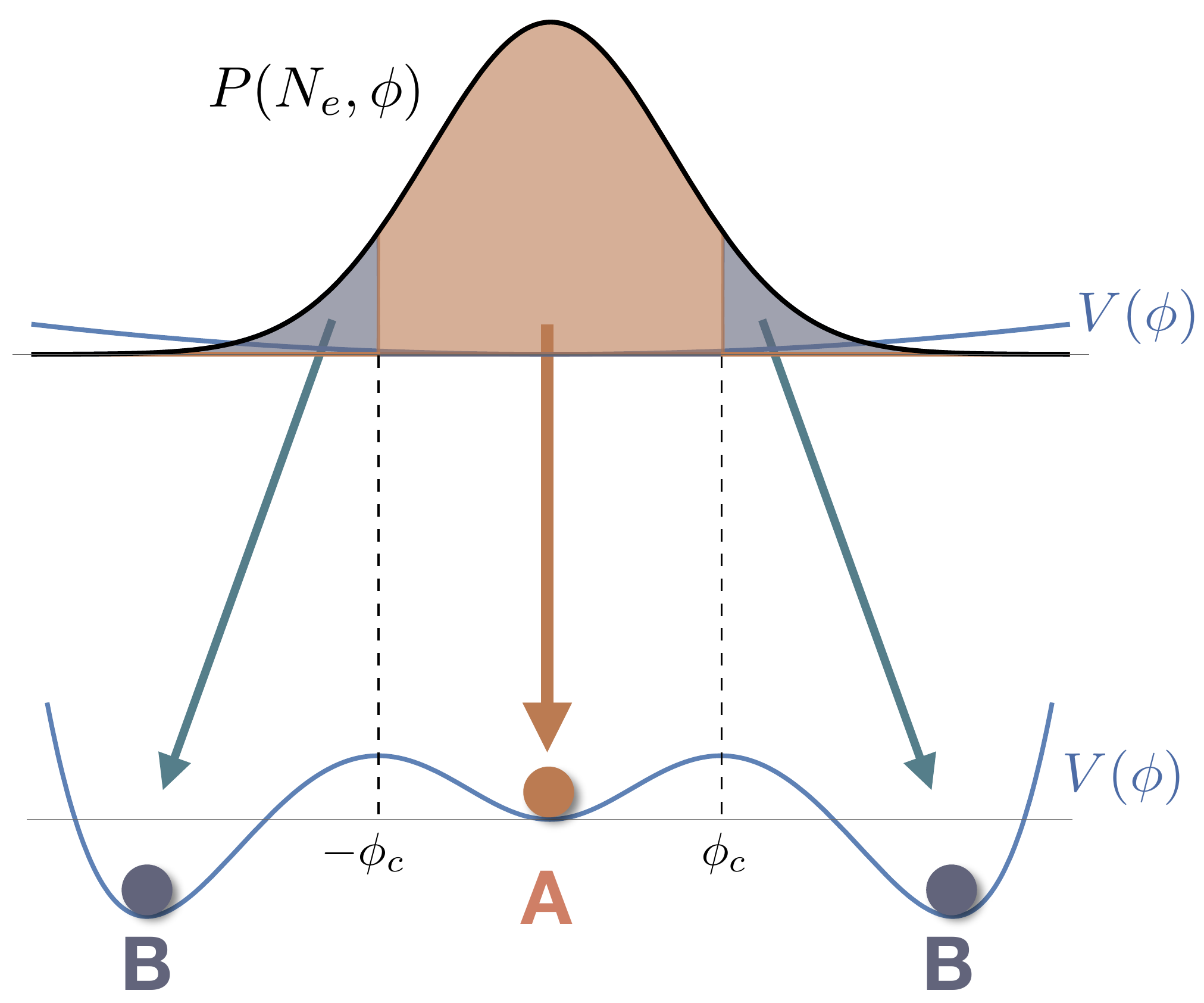}
   \caption{The schematic view of the dynamics of the AD field before and after inflation. {\it Upper side}: During inflation, the IR modes of the AD field diffuse in the complex plain and take different values in different Hubble patches. The probability distribution function $P(N,~\phi)$ is given by the Gaussian form. {\it Lower side}: Just after inflation, the multi-vacua structure appears due to the thermal potential and the negative Hubble induced mass. Some patches where $|\phi|>\phi_c$ is satisfied roll down the vacuum B classically and are identified with HBBs. On the other hand, the others where $|\phi|<\phi_c$ rolls down to the vacuum A.}
   \label{fig:hbb}
\end{figure}

We can see that this setting has a significant feature; \textit{multi-vacua appears after inflation}.
At first, during inflation, the scalar potential has a minimum at the origin ($\varphi=0$) by the positive Hubble induced mass due to the assumption~(i).
After inflation, as usual, the AD field has a vacuum with a non-vanishing VEV [Eq.~(\ref{ADV})], due to the negative Hubble induced mass.
We name this vacuum ``B".
In addition, around the origin $\varphi\lesssim T/c_1$, the mass of the AD field is {\it positive} because the positive thermal mass overcomes the negative Hubble induced mass (= assumption~(ii)), which results in the appearance of the second vacuum $\phi=0$.
We name this new vacuum ``A''.
The shape of the potential after inflation looks like a ``dented" Mexican hat as shown in the lower side of Fig.~\ref{fig:hbb}.
The critical point between two vacuums lies at $\varphi_c(t)\simeq T(t)^2/H(t)$, which is determined by the conditions $V'(\varphi)=0$ and $V''(\varphi)<0$. \\

\subsubsection{Dynamics of the AD field}

Now, let us discuss the dynamics of the AD field in the scenario.
We show the schematic view of the dynamics of the AD field in Fig.~\ref{fig:hbb}.
During inflation, the AD field has positive Hubble induced mass and is located at the origin classically.
However, the AD field acquires quantum fluctuations during inflation.
Therefore, IR modes of the AD field ($=$ coarse-grained AD field over local Hubble patches ) stochastically diffuse in the complex plane and $\phi$ takes different values in different Hubble patches. As we will discuss later, this stage is described by Fokker-Planck equation and the the probability distribution function $P(N,\phi)$ with $N=\ln(a/a_{\rm ini})$ being $e$-folding number, exhibits a Gaussian distribution~\cite{Vilenkin1982,Starobinsky1982,Linde1982}.

After inflation, however, highly non-trivial dynamics occurs.
As we mentioned, the shape of the potential is deformed to the ``dented" Mexican hat and multi-vacua structure appears.
Then, the AD field must roll down to either of the two vacua A and B classically.
If the AD field takes a value $\varphi(t_e)<\varphi_c(t_e)$ in some patch at the end of inflation, $\varphi$ rolls down to the vacuum A.
On the other hand, if $\varphi(t_e)>\varphi_c(t_e)$ is satisfied, $\varphi$ rolls down to the vacuum B.
As a result, the universe is separated into the two phases A and B after inflation. As we will see, although the vacuum B vanishes later, the patches which go through the phase B acquire the baryon asymmetry and form HBBs.

First, we illustrate the evolution of the phase A. After inflation, in the patches with $\varphi(t_e)<\varphi_c(t_e)$, AD fields rolls down and oscillates around the vacuum A. 
In this case, AD field can hardly produce the baryon asymmetry due to the small oscillation amplitude:
\begin{align}
   \eta_b^{({\rm A})}\simeq0
\end{align}
On the other hand, in the patches with $\varphi(t_e)<\varphi_c(t_e)$, AD field rolls down to the vacuum B, that is, AD field makes a condensate as is the case with the conventional AD baryogenesis.
Therefore, at the time $H(t)\simeq H_{\rm osc}$, AD field starts to oscillate toward the vacuum A producing the baryon symmetry given by
\begin{align}\label{etab}
   \eta_b^{({\rm B})}&\simeq
      \epsilon \frac{T_Rm_{3/2}}{H_{\rm osc}^2}
      \left(\frac{\varphi_{\rm osc}}{\Mpl}\right)^2.
\end{align}
Consequently, the separate universe converges to the vacuum A ($\phi=0$) and the difference in their paths in the field space are printed as a difference in the baryon asymmetry.
This is how the inhomogeneous AD baryogenesis works.

In order to apply this mechanism to the creation of the PBH, we assume the produced baryon asymmetry in the phase B is very large such as $\eta_b^{({\rm B})}\sim1$ and the volume fraction of the phase B is very small.
Since outside the HBBs, the substantial baryon asymmetry is not generated, we need another AD field or baryogenesis which realizes the observed homogeneous baryon asymmetry $\eta_b^{\rm ob}\sim10^{-10}$.

Finally we comment on the temperature of the thermal bath produced by the inflaton decay.
To make our analysis, we want to relate the temperature just after inflation $T(t_e)$ to the parameters of the inflation model such as $T_R$ and $ H(t_e)$.
The simplest way is to assume ``instantaneous thermalization" where the evolution of the temperature is give by
\begin{align}
   T^{\rm inst}(t)\simeq(T_R^2H(t)\Mpl)^{1/4}.
\end{align}
Then, since the temperature $T(t)$ decreases slower than $H(t)$, the assumption (ii) is translated as
\begin{align}\label{Deltai}
   \left(\frac{c_1}{c_M}\right)^2\frac{T_R^2\Mpl}{H(t_{e})^3}>1.
\end{align}
However, the actual time scale of the thermalization is not so short~\cite{Mukaida:2015ria}.
As a consequence, the value of $T(t_e)$ could be much lower than the estimation by ``instantaneous thermalization", $T^{\rm inst}(t_e)$.
In our case, however, the required reheating temperature is so high (Eq.~(\ref{Deltai})) that the deviation from the $T^{\rm inst}(t_e)$ is not so significant.\footnote{
See the ref.~\cite{Mukaida:2015ria}.}
Thus assuming $T(t_e)\sim T^{\rm inst}(t_e)$, we rewrite the assumption (ii) as
\begin{align}\label{Delta}
\Delta\equiv \frac{T_R^2\Mpl}{H(t_{e})^3}\gtrsim1,
\end{align}
where we set $c_M,~c_1\sim1$ for simplicity. The critical point just after inflation $\varphi_c(t_e)$  is also rewritten as
\begin{align}
\varphi_c(t_e)\equiv\varphi_c=\Delta^{1/2}H(t_e),
\end{align}
where we also set the $\mathcal{O}(1)$ parameters in the model to unity.

\section{Distribution of HBBs}
\label{sec:HBB_ditribution}

In this section, we evaluate the distribution of the HBBs in terms of the model parameters.
Because the probability distribution function of the AD field during inflation is given by the simple Gaussian form, we can evaluate their distribution analytically.

\subsection{Volume fraction of HBBs}

The evolution of the IR modes of the AD field during inflation is described by stochastic classical theory with the Langevin equation including the Gaussian noise~\cite{Vilenkin1982,Starobinsky1982,Linde1982}.
Then, the evolution of the probability distribution function of the AD field with respect to the $e$-folding number $N$, $P(N,\phi)$ is described by the Fokker-Planck equation as
\begin{align}
   \frac{\partial P(N,\phi)}{\partial N}
      =\sum_{i=1,2} \frac{\partial}{\partial \phi_i}
         \left[
            \frac{V_{\phi_i} P(N,\phi)}{3H^2}
            +\frac{H^2}{8\pi^2}\frac{\partial P(N,\phi)}{\partial \phi_i}
         \right]
\end{align}
where $(\phi_1,~\phi_2)=(\Re[\phi],~\Im[\phi])$.
The first term of the RHS represents the classical force induced by the scalar potential and the second term represents the Gaussian noise due to the quantum fluctuations.
Under the initial condition $P(0,~\phi)\propto \delta(\phi)$ and assuming the Hubble parameter during inflation is constant $H(t\leq t_e)=H_I$, we obtain the analytical expression
\begin{align}\label{qf}
   P(N,\phi)
      =\frac{e^{-\frac{\varphi^2}{2\sigma^2(N)}}}{2\pi\sigma^2(N)},
      ~\ \
   \sigma^2(N)
      =\left(\frac{H_I}{2\pi}\right)^2\frac{1-e^{-c'_IN}}{c'_I}.
\end{align}
Here we have used $V(\phi)\simeq c_IH_I^2\phi^2$ and defined $c'_I\equiv(2/3)c_I$.
The distribution of the phase direction $\theta$ is random unless large CP violating term such as Hubble induced A-terms are introduced.
In the massless limit $c'_I\rightarrow0$, the variance of the AD field growth lineally with respect to $N$ and the scale invariance of the quantum fluctuation is manifest.
For non-vanishing $c'_I$, however, the diffusion of the AD field stops for $N\sim1/c'_I$ due to the classical force induced by the mass term.
Thus, the $c'_I$ determines the saturation of the diffusion.

\begin{figure}[t]
   \centering
   \includegraphics[width=170mm]{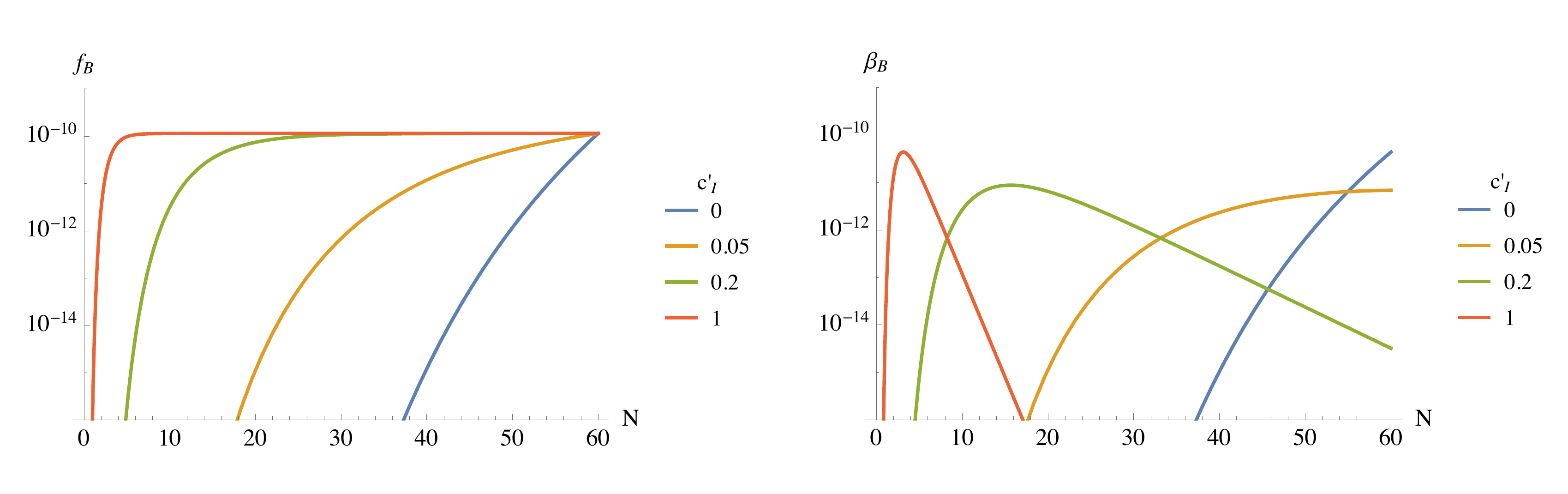}
   \caption{Evolution of the volume fraction of HBBs (left) and production rate of HBBs at $N$ (right) for various $c'_I$.}
   \label{fig:d}
\end{figure}

As we discussed in the previous section, the HBBs are the bubbles which pass through the vacuum B where AD baryogenesis occurs.
Therefore, the physical volume of the HBBs at a certain $N$ is evaluated as
\begin{align}\label{frac}
   V_{\rm B}(N)=V(N)\int_{\varphi>\varphi_c}P(N,\phi)d\phi
   \equiv V(N)f_B(N).
\end{align}
Here we represent the physical volume of the Universe at $N$ as $V(N)\sim r_H^3e^{3N}$, where $r_H$ is the Hubble radius during inflation ($\sim H_I^{-1}$).
$f_B(N)$ represents the volume fraction of the HBBs at $N$.
Since the integration of $f_B$ is straightforward, we can obtained the analytic expression for $f_B$ such that
\begin{align}
   f_B(N)&=
      \int_0^{2\pi}d\theta\int^\infty_{\varphi_c}\varphi
      \frac{e^{-\frac{\varphi^2}{2\sigma^2(N)}}}{2\pi\sigma^2(N)}d\varphi
      =e^{-\frac{2\pi^2\Delta}{\tilde{\sigma}^2(N)}},
\end{align}
where we define $\tilde{\sigma}^2(N)\equiv(1-e^{-c'_IN})/c'_I $.
It is seen that $\Delta$ is responsible for the overall magnitude of the HBB fraction.
We show the evolution of $f_B(N)$ in Fig.~\ref{fig:d} (left panel).
In the figure we take different $\Delta$ for each $c'_I$ fixing the consequential fraction of the HBBs $f_B(N_e)$.
For the larger value of $c'_I$, the growth of $f_B$ saturates earlier because the larger mass ($\propto c'_I$) suppress the quantum diffusion of the AD field.

\subsection{Size distribution of HBBs}\label{3B}

The creation rate of the HBBs at $N$ is understood by differentiating $V_B(N)$ with respect to $N$:
\begin{align}
   \frac{dV_{\rm B}(N)}{dN}
   =3V_{\rm B}(N)
   +V(N)\int_{\varphi>\varphi_c}\frac{dP(N,\phi)}{dN}d\phi.
\end{align}
We can see that the first term in the RHS represents the growth of the existing HBBs due to the cosmic expansion.
The second term represents nothing but the creation of the HBBs at $N$.
Since the created HBBs also grow by the cosmic expansion, the fraction of the HBBs formed at $N$ evaluated at the inflation end $N_e$ is
\begin{align}\label{beta}
   \beta_B(N;N_e)
      =\frac{1}{V(N_e)}\cdot e^{3(N-N_e)}\cdot
      \left[V(N)\int_{\varphi>\varphi_c}\frac{dP(N,\phi)}{dN}d\phi\right].
\end{align}
The third term in the RHS represents the physical volume of the HBBs produced at $N$ and the second term represents their growth due to the cosmic expansion.
Since only the cosmic expansion can not change the volume fraction of the HBBs, the expression of Eq.(\ref{beta}) is actually independent of $N_e$ and reduced to
\begin{align}
   \beta_B(N)=\frac{d}{dN}f_B(N)~
   \left(=\frac{(\pi c'_I)^2\Delta}{\sinh(c'_IN/2)}f_B(N)\right),
\end{align}
as expected.
This quantity is nothing but the volume-fraction distribution of the HBBs created at $N$.
We show the shape of $\beta_B(N)$ in Fig.\ref{fig:d} (right panel).
Indeed, the peak of the HBB distribution coincides with the time when the growth of $f_B(N)$ saturates.
We can see for much smaller $c'_I$, the AD field does not feel the ``viscosity" by the mass term and the production of the  HBBs does not saturate before the end of inflation.
In conclusion, the distribution of the HBBs can be represented only by $c_I$ and the parameter $\Delta$ defined in Eq.(\ref{Delta}).

For later convenience, we relate the size of the HBBs to the horizon mass $M_{\rm H}$ evaluated at the time when they reenter the horizon.
Since the scale of the HBBs are same as the horizon when they are created, we can rewrite the scale of HBBs $k$ created at $N$ as
\begin{align}
   k(N)=k_*e^{N-N_{\rm CMB}},
\end{align}
where $k_*$ is the CMB pivot scale and $N_{\rm CMB}$ is the number of $e$-foldings when the pivot scale exits the horizon.
Since the horizon mass when the scale $k$ reenter the horizon is evaluated as
\begin{align}
   M_{\rm H}
      \simeq19.3M_\odot\left(\frac{g_*}{10.75}\right)^{-1/6}
      \left(\frac{k}{{10^6\rm Mpc}^{-1}}\right)^{-2},
\end{align}
the number of $e$-foldings $N$ is represented in terms $M_{\rm H}$ as
\begin{align}
   N(M_{\rm H})
   \simeq-\frac{1}{2}\ln\frac{M_{\rm H}}{M_\odot}+21.5+N_{\rm CMB},
\end{align}
where $M_\odot$ is the solar mass and we used $g_*=10.75$ and the CMB pivot scale $k_*=0.002{\rm Mpc}^{-1}$.
It is known that the typical inflation models suggest that
\begin{align}
   N_e-N_{\rm CMB}\sim50-60,
\end{align}
On the other hand, to solve the horizon and flatness problem of the Big Bang cosmology, total number of $e$-foldings of the inflation era should be
\begin{align}
   N_e-N_{\rm ini}=N_e\gtrsim60.
\end{align}
Although these values are depend on the post-inflationary thermal history, in the rest of the paper, except for Sec.\ref{sec:double_inflation}, we fix $N_e=60$ and parametrize $N_{\rm CMB}$ as
\begin{align}
   N_{\rm CMB}\sim0-10.
\end{align}

It is also convenient to relate the temperature $T$ to the horizon mass at the horizon reentry as
\begin{align}
   T(M_{\rm H})
      &=434{\rm MeV}\left(\frac{M_{\rm H}}{M_\odot}\right)^{-1/2},
      \\\label{TM}
   M_{\rm H}(T)
      &=18.8M_\odot\left(\frac{T}{200{\rm MeV}}\right)^{-2}.
\end{align}

\section{Gravitational collapse of HBBs}
\label{sec:gravitatinal_collapse}

The purpose of this section is to clarify the thermal history inside the HBBs and discuss their gravitational collapse into PBHs.
The energy density inside and outside HBBs are almost same just after inflation because of the energy conservation of the AD field and the domination of the oscillation energy of the inflaton.
However, after AD baryogenesis takes place, the energy contrast between outside and inside the HBBs becomes significant resulting in the formation of the PBHs.

\subsection{Density contrast of HBBs}

As we discussed in the previous sections, the HBBs are naturally produced in the consequence of the AD baryogenesis.
Inside the HBBs, we assume that there is large baryon asymmetry such as $\eta_b^{\rm in}\simeq\eta_b^{\rm (B)}\sim 1$, while outside the HBBs the baryon asymmetry is $\eta_b^{\rm out}=\eta_b^{\rm ob}\sim10^{-10}$, which must be realized by additional baryogenesis.
This fluctuation of the baryon number density is regarded as a top-hat type isocurvature perturbation.
Such a large but small-scale isocurvature perturbation is hardly constrained by the observations in contrast to the adiabatic perturbations.
That is why this model does not suffer from the stringent constraints from CMB $\mu$-distortion and the PTA experiment.
In the followings we see that the isocurvature perturbations due to HBBs induce substantial density perturbations.

\subsubsection{QCD phase transition}

After the AD field decays, the produced baryon number is carried by the quarks.
As long as the quarks remain relativistic particles, the density fluctuations are not produced.
However, nearly massless quarks are confined within the massive baryons ($=$protons and neutrons) by the QCD phase transition.
Thus, after the QCD phase transition the energy of the baryons behave as matter and their energy density is $\rho\simeq n_b m_b $, where $m_b$ is the nucleon mass.
The point is that inside the HBBs, the contribution of the baryons to the total energy density is large while it is negligible outside the HBB.
The density contrast between inside and outside the HBBs is represented as
\begin{align}\label{T}
   \delta\equiv\frac{\rho^{\rm in}-\rho^{\rm out}}{\rho^{\rm out}}
   \simeq\frac{n_b^{\rm in} m_b}{(\pi^2/30)g_*T^4}
   \simeq0.3\eta^{\rm in}_b
      \left(\frac{T}{200{\rm MeV}}\right)^{-1}
      \theta(T_{\rm QCD}-T),
\end{align}
where we have used $m_b\simeq938{\rm MeV}$ and $\theta(x)$ is the Heaviside theta function.
This value can be larger enough to form the PBHs for $\eta_b^{\rm in}\sim1$.
Strictly speaking, the temperature of the plasma inside the HBBs is different from that outside the HBBs due to the large chemical potential.
However, even in our case $\eta_b^{\rm in}\sim1$, the deviation is at most $\mathcal{O}(1)$ and hence we neglect the difference.

Here we make a comment on the fluctuations of the baryon asymmetry $\eta^{\rm in}_b$ among HBBs.
The baryon asymmetry produced by the AD mechanism depends on the initial value of the phase direction of the AD field $\theta_0$ [Eq.(\ref{CAD})].
Since the phase direction $\theta$ is almost flat during inflation, that is, there in no CP-violating terms other than the tiny soft A-term, the values in the HBBs of $\theta_0$ are generally different.
Thus the baryon asymmetry inside the HBBs are different by factor $\sin(\theta_0-\arg(a_M))$, which clearly affects the evaluation of the density contrast Eq.(\ref{T}).
Although this effect does not bring a substantial change to the following discussion, we can realize the uniform baryon asymmetry by introducing the Hubble induced A-term.

\subsubsection{Q-ball formation}

We have just considered that the coherent oscillation of the AD field decays into the quarks after the $t>t_{\rm osc}$.
However, it is known that the coherent oscillation of the AD field is usually spatially unstable and fragments to the localized lumps, called Q-balls.
A Q-ball is a configuration of the complex scalar which minimizes the energy under the fixed global $U(1)$ charge (in this case, which is identified with a baryon number).
The property of the Q-balls formed after AD baryogenesis depends on the scenario of the mediation of the SUSY breaking effect. In the gravity-mediated SUSY breaking scenario, produced Q-balls are unstable against the decay to the quarks.
Therefore, the baryon number is produced from their decay and hence the Q-balls are only the ``transients" of the baryogenesis.
On the other hand, in the gauge-mediated SUSY breaking scenario, Q-balls are stable and behave as the dark matter~\cite{Kasuya2001,Kasuya2000g}.
We represent the abundance of the stable Q-balls formed after the AD baryogenesis inside the HBBs as $Y^{\rm in}_Q\equiv\rho_Q^{\rm in}/s$.
Since Q-balls are not produced outside the HBBs, the density contrast becomes
\begin{align}\label{delta}
   \delta
      =\frac{\rho_Q^{\rm in}}{(\pi^2/30)g_*T^4}
      =\frac{4}{3}\left(\frac{T}{Y^{\rm in}_Q}\right)^{-1},
\end{align}
which can clearly reach $\mathcal{O}(1)$ at $T\sim Y^{\rm in}_Q$.
It is known that all the baryon number produced by the AD baryogenesis is enclosed in the Q-balls and their abundance $Y^{\rm in}_Q$ is evaluated by the Q-ball number density and the Q-ball mass.

\subsection{PBH formation}

As we have seen, the HBBs become over-dense by the QCD phase transition or the formation of stable Q-balls.
If the density contrast is large enough, the self-gravity of the over-dense regions overcomes the pressure and the regions gravitationally collapse into PBHs just after the horizon reentry.
In the radiation-dominated era, the threshold value of the density contrast for the PBH formation is estimated as $\delta_c\simeq w$~\cite{Carr:1974nx}, where $w\equiv p/\rho$ is the parameter of the equation of the state.
Although the recent studies perform the more precise estimation of $\delta_c$ by analytic/numerical method, we simply adopt $\delta_c\simeq w$ because our model is hardly sensitive to the choice of the $\delta_c$ unlike the case of the Gaussian density perturbations.

Then let us consider the PBH formation in this model according to the threshold value $\delta_c\simeq w$.
The characteristic point of this model is that density perturbations do not conserve even in super-horizon scale due to the redshift of the radiation.
The condition for the PBH formation is represented by
\begin{align}
   \delta(T)> \delta_c(T)\simeq w(T)
\end{align}
and it depends on the temperature at the horizon crossing of the HBB.
Since the density contrast of the HBB is originated from the non-relativistic (pressure-less) baryons/Q-balls, the equation of the state parameter $w$ inside the HBB can be expressed in terms of $\delta(T)$ as
\begin{align}
   w(T)
      =\frac{p^{\rm in}}{\rho^{\rm in}}
      \simeq\frac{p^{\rm out}}{\rho^{\rm in}}
      =\frac{1}{3}\frac{1}{1+\delta(T)}.
\end{align}
Therefore, the condition of the PBH formation is written as
\begin{align}
   \delta(T) \gtrsim \frac{1}{3}\frac{1}{1+\delta(T)}
   ~\Longleftrightarrow~
   \delta(T) \gtrsim 0.26,
\end{align}
and this gives an upper bound on the temperature at the horizon crossing of the HBB.
This critical temperature for the PBH formation $T_c$ is obtained from Eq.(\ref{T}) and (\ref{delta}) such that
\begin{align}
   T_c\simeq
   \begin{cases}
      \Min[231\eta_b^{\rm in}{\rm MeV},~T_{\rm QCD}],
      &(\text{QCD phase transition}) \\
      5.1Y_Q^{\rm in}.
      & (\text{stable Q-ball~formation})
   \end{cases}
\end{align}
According to Eq.(\ref{TM}), these conditions are translated to the lower bound on the horizon mass at the horizon reentry.
This critical value of the horizon mass for PBH formation is then given by
\begin{align}
   M_c\simeq
   \begin{cases}
      \Max\left[14.1(\eta_b^{\rm in})^{-2}M_\odot,
         ~18.8M_\odot\left(\frac{T_{\rm QCD}}{200{\rm MeV}}\right)^{-2}\right],
      &(\text{QCD phase transition}) \\
      18.1M_\odot\left(\frac{Y_Q^{\rm in}}{40{\rm MeV}}\right)^{-2}.
      & (\text{stable Q-ball~formation})
  \end{cases}
  \label{eq:critical_mass}
\end{align}
Thus, only the HBBs larger than $M_c$ can gravitationally collapse into the PBHs.
On the other hand, smaller HBBs can not collapse due to the pressure inside the HBBs, but would form the self-gravitating objects made of baryons/Q-balls.
The interesting point is that the mass distribution of the PBHs is determined by not only the distribution of the HBBs, but also this cutoff $M_c$.
Assuming the formed PBH has the mass which is comparable with the horizon mass at the reentry, $M_{\rm PBH}\sim M_{\rm H}$, we can evaluate the distribution of the PBHs in the model as
\begin{align}
   \label{fr}
   \beta_{\rm PBH}(M_{\rm PBH})=\beta_B(M_{\rm PBH})\theta(M_{\rm PBH}-M_c).
\end{align}
In the case of the stable Q-ball formation, the cutoff scale of the PBH formation is determined by the Q-ball abundance generated by the AD baryogenesis. However, in the case without the stable Q-ball formation, the cutoff scale must be $\mathcal{O}(10)M_\odot$, that is, mass range of the BHs detected by LIGO is naturally explained. This is the most fascinating feature of the model because any fine-tuning of the parameters is required unless $\eta^{\rm in}_b\sim1$ is realized.

Here we make a comment on the PBH formation at $T\ll T_c$.
Since the density contrast $\delta$ grows linearly in time due to the redshift, we naively expect $\delta$ becomes larger than unity at a certain time.
It had been pointed out such a large density contrast indicates a separate universe and hence PBH is not formed.
However, the recent study~\cite{Carr:2014pga} suggests the separate universe is not created in realistic cosmological evolution and does not constrain the PBH formation.
Furthermore, even in this case, they show that mass of the created PBH generally can not be much larger than the horizon mass.
Thus, we consider PBH is continued to be formed even when $T\ll T_c$ and their mass is roughly given by the horizon mass $M_{\rm H}$.
The PBH formation from the over-density of the Q-balls has been studied in a different cosmological context \cite{Cotner:2016cvr,Cotner:2017tir}.


\section{LIGO events from Affleck-Dine baryogenesis}
\label{sec:LIGO_event}

Let us discuss the PBH formation from the AD baryogenesis in a concrete model and its consistency with the current observational constraints.
In this section, we consider the gravity-mediated and gauge-mediated SUSY breaking scenarios.
As we mentioned, the difference in the mediator of the SUSY breaking effect is related to the stability of the Q-ball.

\subsection{Gravity-mediated SUSY breaking scenario}

We first consider the case where the SUSY breaking effect is mediated to the visible sector only through gravity.
In this case, taking the one-loop correction into account, the potential of the AD field after the oscillation is written as
\begin{align}
   V(\phi)\simeq m_{3/2}^2|\phi|^2
   \left[1+K\ln\left(\frac{|\phi|^2}{M_*^2}\right)\right],
\end{align}
where $m_{3/2}$ is the gravitino mass, $M_*$ is a renormalization scale and $K$ is a constant determined by specifying the MSSM flat-direction and typically $-K=0.1\sim0.01$.\footnote{
The gauginos give negative contributions to the one-loop potential while the Yukawa couplings give positive contributions.
In general, the gaugino contribution is dominant and $K$ is negative.
However, $K$ can be positive for flat directions which contain stop.}
Since this one-loop correction makes the potential flatter than the quadratic one, AD field feels the negative pressure and forms localized solitons called Q-balls.
The Q-balls formed by this gravity-mediation potential is called ``gravity-mediation type" and its properties are as follows:
\begin{align}\label{grp}
   M_Q&\simeq m_{3/2}Q,\\
   R_Q&\simeq |K|^{-1/2}m_{3/2}^{-1},\\
   \omega_Q&\simeq m_{3/2},
\end{align}
where $M_Q$ and $R_Q$ are the mass and the size of the Q-ball and $\omega_Q$ is the energy of the Q-ball per unit baryon number.
The gravity-mediation type Q-ball is unstable with respect to the decay into the nucleons.
This is simply because $\omega\simeq m_{3/2}$, which can be regarded as the effective mass of the AD field, is larger than the nucleon mass $\simeq 1{\rm GeV}$ in the gravity-mediation where typically $m_{3/2}\gg {\rm GeV}$.
Thus, baryons confined in the Q-balls are released inside the HBBs through the Q-ball decay.
Then, the density contrast of the HBBs are induced by the QCD phase transition as Eq.(\ref{T}) and the critical mass scale of the PBH is $\mathcal{O}(10)M_\odot$ as long as $\eta_b^{\rm in}\sim 1$.

\begin{figure}[t]
   \centering
   \includegraphics[width=180mm]{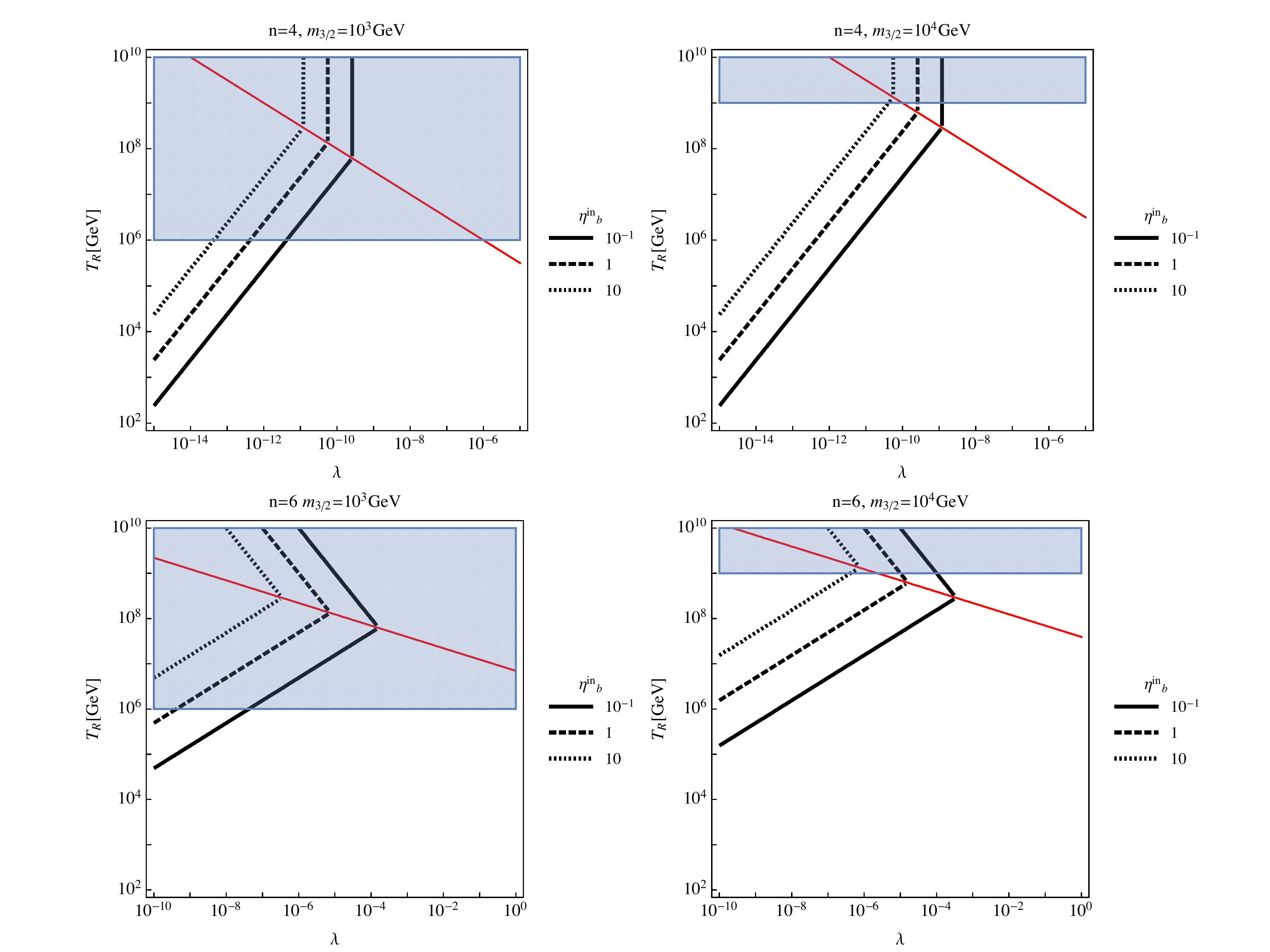}
   \caption{We search the parameter space where the $\mathcal{O}(1)$ baryon asymmetry is produced.
   The upper panels are the case with the $n=4$ AD field and the lower panels are the case with $n=6$.
   The left and right panels correspond to $m_{3/2}=10^3{\rm GeV},10^4{\rm GeV}$ respectively.
   Three black lines represent the contours of $\eta_b^{\rm in}=(10^{-1},1,10)$.
   On the upper side of the red line, the dynamics is dominated by the finite temperature effect.
   The blue shaded region is excluded by the overproduction of the gravitino.}
   \label{fig:gr1}
\end{figure}

Before discussing the abundance of the PBHs, we consider how $\eta_b^{\rm in}\sim 1$ is realized in our model. In fact, AD baryogenesis can naturally produce such huge baryon asymmetry especially in the case of $n=6$. The baryon asymmetry inside the HBB is given by Eq.(\ref{etab}).
Since we are considering the situation where the temperature of the thermal plasma is relatively high due to the assumption (ii), thermal potential Eq.(\ref{TP}) can also trigger the oscillation of the AD field in addition to the soft mass.
Therefore, the Hubble parameter when the AD field starts to oscillate is evaluated as
\begin{align}
   H_{\rm osc}\simeq \Max
   \left[
      m_{\phi},
      ~\Mpl\lambda^{2/n}\left(\frac{T_R}{\Mpl}\right)^{\frac{n-2}{n/2}}
   \right].
\end{align}
Taking this ``early oscillation" into account, we can calculate the consequential baryon asymmetry produces in the HBBs.
In the Fig.~{\ref{fig:gr1}}, we plot the contours of $\eta_b^{\rm in}=(10^{-1},~1,~1)$ in the $(\lambda,~T_R)$ - plane.
In the case of $n=4$, the production of the large baryon asymmetry $\eta_b^{\rm in}\sim1$ requires smaller $\lambda$ such as $10^{-10}$. On the other hand, in the case of $n=6$, the amplitude of the oscillation of the AD field is relatively large and $\lambda\lesssim10^{-5}$ is sufficient.
In addition, we have to take into account the thermal production of the gravitino in both cases.
In the gravity-mediated SUSY breaking scenario, the mass of gravitino tends to heavy ($m_{3/2}\sim 10^{2-3}{\rm GeV}$) and unstable against the radiative/hadronic decay.
It is shown that such decay product could spoil the success of the big bang nucleosynthesis and their abundance must be small.
This fact sets the upper bound on the reheating temperature (blue shaded region in the Fig.~{\ref{fig:gr1}}).\footnote{
Strictly speaking, this upper bound on the reheating temperature has mild dependence on the MSSM parameters such as gaugino mass and scalar mass and so on.
In this paper, we assume the bound is almost independent of the MSSM parameters and use a typical value.
See the ref.~\cite{Kawasaki:2008qe} for detail.}
In any case, the large baryon asymmetry inside the HBBs is consistently produced in this scenario.
In the following, we assume $\eta_b^{\rm in}$ is so large that $M_c=18.8M_\odot(T_{\rm QCD}/200\text{MeV})^{-2}$ is realized.

 \begin{figure}[t]
   \centering
   \includegraphics[width=128mm]{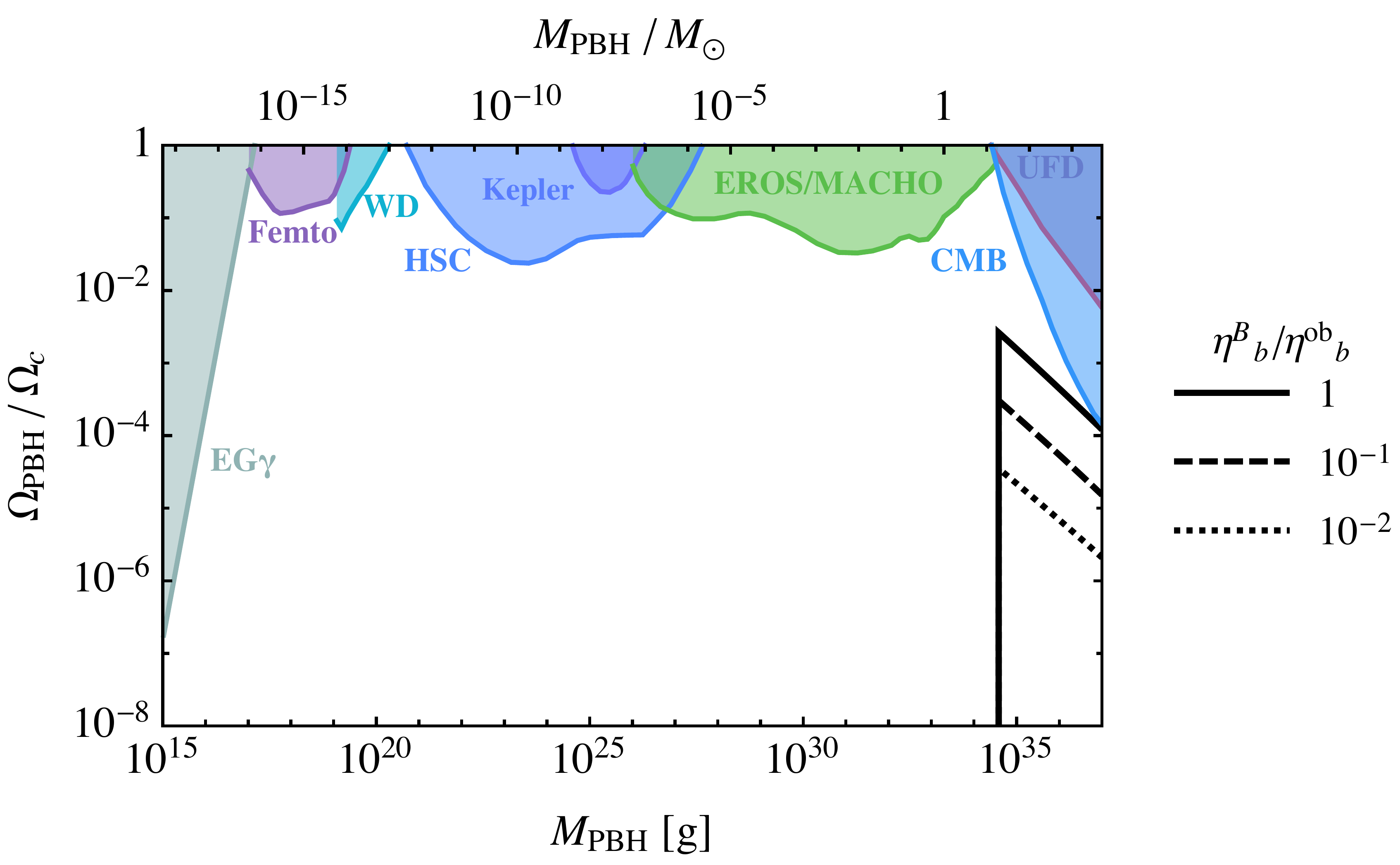}
   \caption{We show the PBH abundance and the observational constraints.
       The shaded regions are excluded by extragalactic gamma rays from Hawking radiation (EG$\gamma$)~\cite{Carr2009}, femtolensing of known gamma ray bursts (Femto)~\cite{Barnacka2012}, white dwarfs existing in our local galaxy (WD)~\cite{Graham2015}, microlensing search with Subaru Hyper Suprime-Cam (HSC)~\cite{Niikura2017}, Kepler micro/millilensing (Kepler)~\cite{Griest2013}, EROS/MACHO microlensing (EROS/MACHO)~\cite{Tisserand2007}, dynamical heating of ultra faint dwarf galaxies (UFD)~\cite{Brandt2016}, and accretion constraints from CMB (CMB)~\cite{Ali-Haimoud20172}.
}
   \label{fig:qcd}
\end{figure}

Then, let us discuss the abundance of the PBHs.
The present abundance of the PBHs with mass $M_{\rm PBH}$ over logarithmic mass interval $d(\ln M_{\rm PBH})$ are estimated as
\begin{align}
   \frac{\Omega_{\rm PBH}(M_{\rm PBH})}{\Omega_c}
      &\simeq\left.\frac{\rho_{\rm PBH}}{\rho_m}\right|_{\rm eq}\frac{\Omega_m}{\Omega_c}
         =\frac{\Omega_m}{\Omega_c}\frac{T(M_{\rm PBH})}{T_{\rm eq}}
         \beta_{\rm PBH}(M_{\rm PBH})\\\label{pa}
      &\simeq\left(\frac{\beta_{\rm PBH}(M_{\rm PBH})}{1.6\times10^{-9}}\right)
         \left(\frac{\Omega_ch^2}{0.12}\right)^{-1}
         \left(\frac{M_{\rm PBH}}{M_\odot}\right)^{-1/2},
\end{align}
where $\Omega_c$ and $\Omega_m$ are the present density parameters of the dark matter and matter, respectively.
Here we use the latest Planck result $\Omega_ch^2\simeq0.12$~\cite{PlanckCollaboration2015a} [$h$: the present Hubble parameter in units of $100\,\text{km/sec/Mpc}$].
$T(M_{\rm PBH})$ and $T_{\rm eq}$ are the temperatures at the formation of the PBHs with mass $M_{\rm PBH}$ and the matter-radiation equality, respectively.

The other parameters which determine $\beta_{\rm PBH}$ are $c_I'$, $\Delta$ and $N_{\rm CMB}$.
In order to evaluate not only the created PBHs but also the effect of the residual HBBs, we introduce the quantity
\begin{align}
   \eta_b^{\rm B}\equiv f_B(N_e)\eta_b^{\rm in},
\end{align}
which represents the contribution of the HBBs to the baryon asymmetry of the entire universe.
We show the prediction of the PBH abundance for $\eta_b^{\rm B}/\eta_b^{\rm ob}=(1,10^{-1},10^{-2})$ in Fig.~\ref{fig:qcd}.
We can see that due to the cut-off $M_c$, there exists a peak-like ``edge" whose mass $\sim\mathcal{O}(10)M_\odot$.
The figure confirm the fact that large $\eta_b^{\rm B}$ realize the higher peak.
The LIGO events can be explained for $\mathcal{O}(10)M_\odot$ PBHs whose abundance is $\Omega_\text{PBH}/\Omega_c \sim \mathcal{O}(10^{-3})$--$\mathcal{O}(10^{-2})$~\cite{Sasaki2016}, which requires $\eta_b^{\text{B}}\sim \eta_b^{\text{ob}}$.
However, $\eta_b^{\rm B}$ can not exceed the observational baryon asymmetry $\eta_b^{\rm ob}\sim10^{-10}$.
Furthermore, since the baryon number density inside the HBB, $\eta_b^{\rm in}$, is so huge that the produced abundances of the light elements are significantly different from the prediction of ordinary BBN.
Thus, in order not to spoil the success of BBN, $\eta_b^{\rm B}/\eta_b^{\rm ob}\ll1$ is required.
Although the model still have the sizable contribution to the LIGO event, the abundance may be too small to account for all LIGO events.
As we will see later, a double inflation scenario enable us to obtain more higher and sharper peak even if $\eta_b^{\rm B}/\eta_b^{\rm ob}\ll1$ is satisfied.

We note on the possible form of the residual HBBs in the present time.
Since the density contrast of the HBBs reaches the order of unity after QCD phase transition, we naively expect that they form a self-gravitating system of the non-relativistic baryons where a significant amount of heavy elements may be synthesized by BBN.
There is a possibility that they contaminate the surrounding universe and explain the metallicity of the population II stars.
On the other hand, it is known that QCD phase transition may give birth to the hypothetical bound state of up, down and strange quarks, called quark nuggets (or strange matters)\cite{Witten:1984rs, Alcock:1985vc, Iso:1985iw}.
Although there have been some research on them because they are possible candidate for baryonic dark matter, it is revealed that almost all of the quark nuggets evaporate and can not survive until now.
For the quark nuggets to survive against the evaporation, they have to carry a large baryon number such as $N_B\gtrsim10^{51}$, which is much larger than the baryon number in the horizon at QCD epoch.
In the our scenario, however, the formation of the stable quark nuggets may possible.
This is simply because HBBs can carry anomalously large baryon number greater than $10^{51}$.
In fact, the HBBs with mass greater than $10^{-5}M_{\odot}$ would form stable quark nuggets and contribute to the current dark matter.
While their abundance is negligible with respect to the entire abundance of the HBBs, they may reduce the contribution of the HBBs to the net baryon asymmetry, $\eta_b^{\rm B}$, and relax the constraint on the PBH abundance.

\subsection{Gauge-mediated SUSY breaking scenario}

 \begin{figure}[t]
   \centering
   \includegraphics[width=180mm]{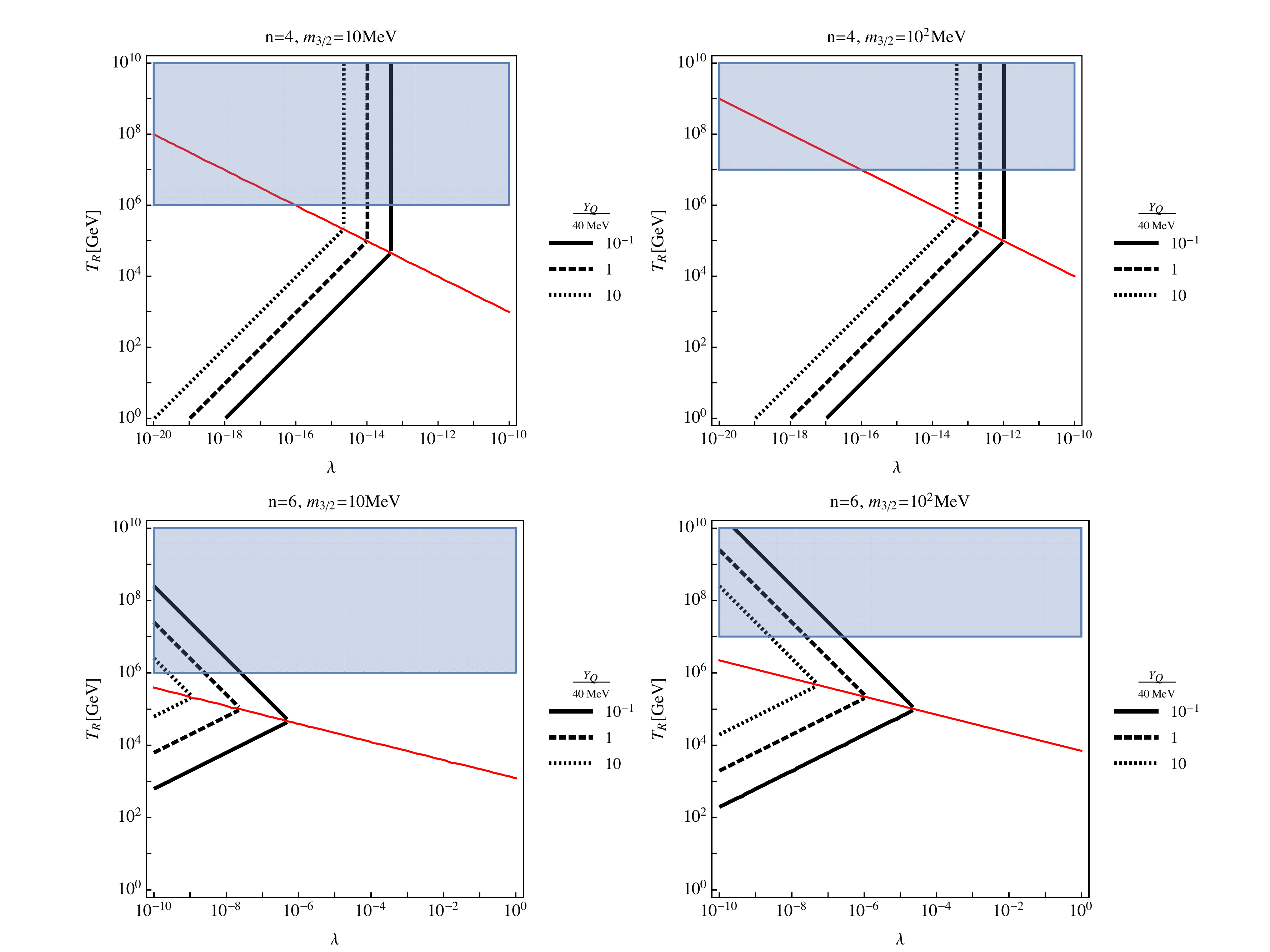}
   \caption{We search the parameter space where the $Y_{Q}^{\rm in}\sim 40{\rm MeV}$ is realized.
   The upper panels are the case with the $n=4$ AD field and the lower panels are the case with $n=6$.
   The left and right panels correspond to $m_{3/2}=10{\rm MeV},10^2{\rm MeV}$ respectively.
   Three black lines represent the contours of $Y_{Q}^{\rm in}/40{\rm MeV}=(10^{-1},1,10)$.
   On the upper side of the red line, the dynamics is dominated by the finite temperature effect.
   The blue shaded region is excluded by the overproduction of the gravitino which exceeds the current dark matter abundance.}
   \label{fig:Q1}
\end{figure}

Next, we consider the case of the gauge-mediated SUSY breaking.
In this case, the potential of the AD field is lifted up as
\begin{align}
   V(\phi)\simeq
       M_{\rm F}^4\left(\log\frac{|\phi|^2}{M_{\rm mess}^2}\right)^2
       +m_{3/2}^2|\phi|^2\left[1+K\ln\left(\frac{|\phi|^2}{M_*^2}\right)\right].
\end{align}
Here we include the contribution from the gravity-mediation (2nd term) because it is not forbidden generically.
We can see that the contribution from the gauge-mediation (1st term) is also shallower than the quadratic potential and the AD field fragments to the Q-balls.
Since there are two contributions to the potential, two types of the Q-balls exist in the gauge-mediation, ``gauge-mediation type" and ``new type".
The former is realized when the potential is dominated by the gauge-mediation potential and the Q-ball properties are
\begin{align}
   M_Q&\simeq \frac{4\sqrt{2}\pi}{3}\zeta M_{\rm F}Q^{3/4}\\
   R_Q&\simeq \frac{1}{\sqrt{2}}\zeta^{-1} M_{\rm F}^{-1}Q^{1/4},\\
   \omega_Q&\simeq \sqrt{2}\pi\zeta M_{\rm F}Q^{-1/4},
\end{align}
where $\zeta$ is a $\mathcal{O}(1)$ numerical constant.
We can see this Q-ball configuration is stable against the decay into the nucleons for sufficiently large charge.
On the other hand, the later is realized when the potential is dominated by the gravity-mediation potential and the Q-ball properties are given by
\begin{align}
   M_Q&\simeq m_{3/2}Q,\\
   R_Q&\simeq |K|^{-1/2}m_{3/2}^{-1},\\
   \omega_Q&\simeq m_{3/2},
\end{align}
which are the same as those of the ``gravity-mediation type" (Eq.(\ref{grp})) because the scalar potential is identical.
However, in the gauge-mediation scenario, the gravitino mass $m_{3/2}$ is typically lighter than $\sim {\rm GeV}$ and the decay to the nucleons are forbidden by the kinematics.
Thus, the gauge-mediation scenario ensures the existence of the stable Q-balls which are good candidate of the dark matter.
Since the gauge-mediation potential also triggers the oscillation of the AD field, the Hubble parameter at the start of the oscillation is given by
\begin{align}
   H_{\rm osc}\simeq \Max
   \left[
       m_{3/2},
       ~\left(\lambda\frac{M_{\rm F}^{2n-4}}{\Mpl^{n-3}}\right)^{\frac{1}{n-1}},
       ~\Mpl\lambda^{2/n}\left(\frac{T_R}{\Mpl}\right)^{\frac{n-2}{n/2}}
   \right],
\end{align}
which determines the total baryon asymmetry enclosed in the Q-balls.
The type of the Q-ball is specified by comparing $\phi_{\rm osc}$ with the critical point $\phi_{\rm eq}\simeq M_{\rm F}^2/m_{3/2}$ where gravity- and gauge- mediation potential become comparable. The new (gauge-mediation) type Q-ball is obtained for $\phi_{\rm　osc}>\phi_{\rm eq}~(\phi_{\rm osc}<\phi_{\rm eq})$.
In the aim of the creation of the PBH, however, the required baryon number density is so high that the contribution from the gauge mediation potential is negligible.
Thus the produce Q-balls are determined to the new type ones and their abundance is estimated as
\begin{align}\label{qad}
   Y_{Q}^{\rm in}
      =M_Q\frac{n_Q^{\rm in}}{s}
      =\frac{M_Q}{Q}\eta_b^{\rm in}
      \simeq　m_{3/2}\eta_b^{\rm in}.
\end{align}
From now on, restricting our interest to the PBH with mass $\mathcal{O}(10)M_\odot$, let us consider the case where the Q-ball abundance satisfies $Y_{Q}^{\rm in}\sim 40{\rm MeV}$ [see, Eq.~(\ref{eq:critical_mass})].
We calculate the Eq.~(\ref{qad}) and plot the contours of $Y_{Q}^{\rm in}/40{\rm MeV}=(10^{-1},~1,~10)$ in the $(\lambda,~T_R)$ - plane in Fig.~\ref{fig:Q1}.
As with the case in the gravity-mediated SUSY breaking scenario, there exist a parameters which realize the sufficient amount of the Q-balls.
Here we also make a remark about the thermally produced gravitino.
In the gauge-mediated SUSY breaking scenario, gravitino is the lightest SUSY particle (LSP) and they contribute to the dark matter abundance.
Therefore, the reheating temperature must have a upper bound (blue shaded region in the Fig.~\ref{fig:Q1}) for the thermally produced gravitino not to over-close the universe~\cite{Moroi:1993mb}, which is stronger than the case of the gravity-mediation.

 \begin{figure}[t]
   \centering
   \includegraphics[width=110mm]{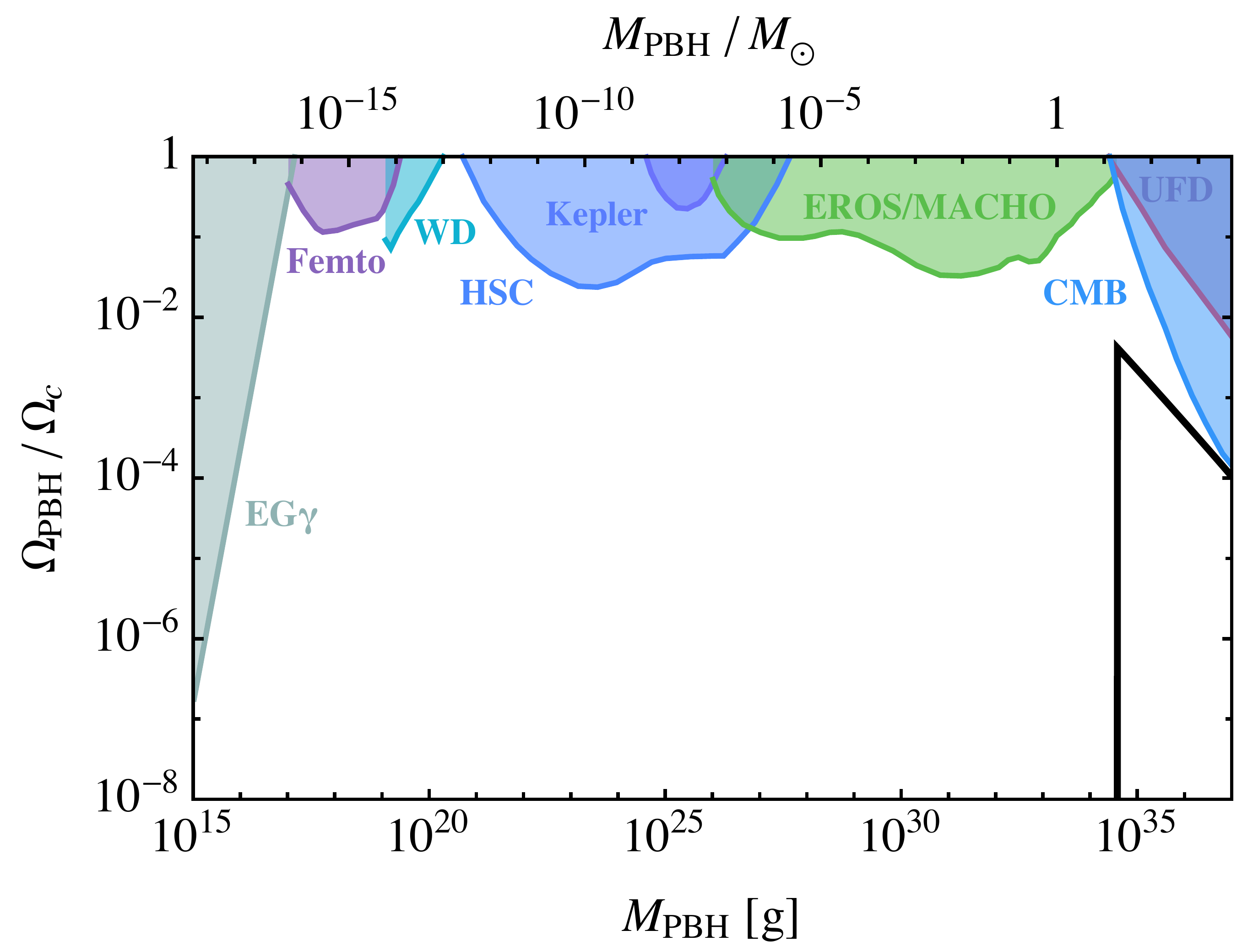}
   \caption{We show the PBH abundance in the case of the gauge-mediated SUSY breaking scenario, where PBHs are formed by the over-density of the Q-balls. Here we make the parameter choice $(c_I,\Delta,N_{\rm CMB})=(0.046,19,10)$. The observational constraint are represented by the shaded region by the same manner with that in Fig.~\ref{fig:qcd}.
}
\label{fig:q}
\end{figure}

\begin{figure}[t]
   \centering
   \includegraphics[width=100mm]{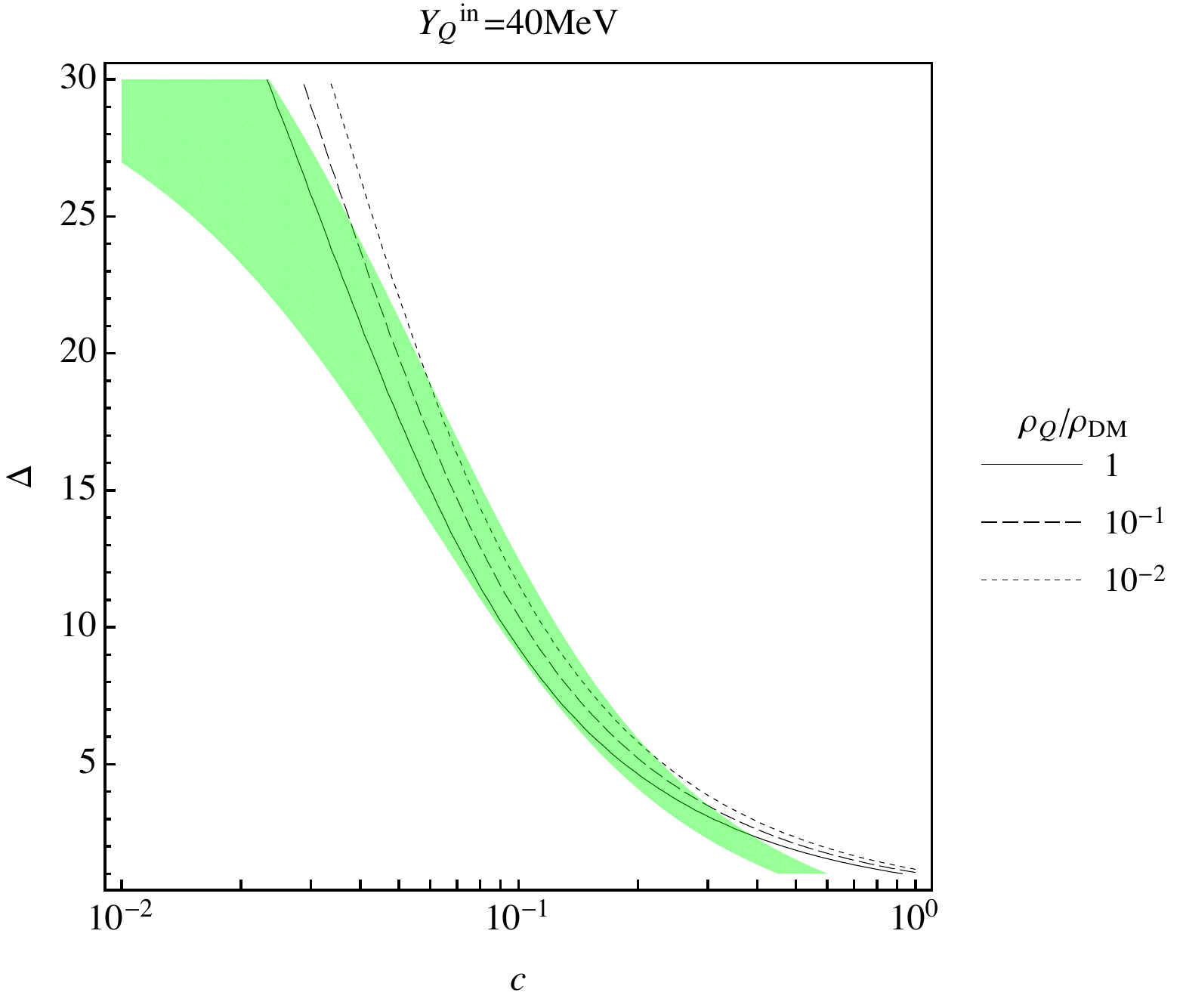}
   \caption{The green shaded region is where the peak of the PBH abundance is consistent with the event rate inferred from the LIGO event, $\Omega_{\rm PBH}/\Omega_c\sim\mathcal{O}(10^{-2})-\mathcal{O}(10^{-3})$.
   We also show the region where the energy density of the Q-balls could give a sizable contribution to the current dark matter density by black lines.
   It is found that two regions are well degenerated in the almost all of the $(c_I,\Delta)$-plane.}
\label{fig:cg}
\end{figure}

Next, let us discuss the abundance of the PBHs in this case. Using Eq.(\ref{pa}) again we show the one example which can explain the LIGO event evading the observational constraints in Fig.~\ref{fig:q}.
In contrast to the case without Q-balls, the peak at the $M_c$ is so high that the event rate $\Omega_{\rm PBH}/\Omega_c\sim\mathcal{O}(10^{-2})-\mathcal{O}(10^{-3})$~\cite{Sasaki2016} is easily realized.
This is simply because the residual HBBs, which is too small to collapse into the PBHs, do not contribute to the baryon asymmetry and larger HBBs abundance such as $f_B\sim10^{-8}$ is permitted.
On the other hand, the Q-balls inside the residual HBBs would survive until now and contribute to the current dark matter abundance.
This contribution is estimated as
\begin{align}\nonumber
   \frac{\rho_Q}{s}&\simeq f_BY_Q^{\rm in}
   =4.4\times10^{-10}{\rm GeV}
      \left(\frac{Y_Q^{\rm in}}{40{\rm MeV}}\right)
      \left(\frac{f_B}{1.1\times10^{-8}}\right).
\end{align}
Interestingly, the value of $f_B$ required to realize the LIGO event rate is very similar to one which make the residual Q-balls constitute to the all dark matter.
We plot the region where the event rate of the BH merger $\Omega_{\rm PBH}/\Omega_c\sim\mathcal{O}(10^{-2})-\mathcal{O}(10^{-3})$ is explained and the contribution of the residual Q-balls to the current dark matter abundance in the $(c_I,\Delta)$-plane in Fig.~\ref{fig:cg}.
From this figure, we can conclude that the residual Q-balls inevitably make a sizable contribute to the dark matter abundance to explain the LIGO event rate.
In other words, the LIGO PBHs and dark matter are simultaneously generated, namely, {\it cogenerated} in our scenario.
Actually, the parameter choice $(c_I,\Delta,N_{\rm CMB})=(0.046,19,10)$ we made in the Fig.~\ref{fig:q} explains the all dark matter by the residual Q-balls.

Before closing this section, we make a comment on the scale of the inflation in the gauge-mediated SUSY breaking scenario.
We have seen that the reheating temperature has a upper bound $10^{6-7}{\rm GeV}$ for gravitino not to over-close the universe, that is, $Y_{3/2}\leq Y_{\rm DM}$.
However, this requirement is not sufficient because residual Q-balls must have a significant contribution to the dark matter.
Therefore, the reheating temperature must be much lower than $10^{6-7}{\rm GeV}$.
In this case, the condition for domination of the thermal effect over the Hubble induced mass [Eq.~(\ref{Delta})] is satisfied only if the Hubble parameter during inflation is small ($H_I \lesssim 10^{10}$~GeV).
As a result, this scenario works only in the low-scale inflation scenario such as new-inflation ($H_I \sim 10^{6}$~GeV), $\alpha$-attractor inflation with small $\alpha$ ($H_I \sim \sqrt{\alpha}10^{13}$~GeV), and some string-motivated models with small tenser-to-scalar ratio $r$.

\section{Extension to the double inflation scenario}
\label{sec:double_inflation}

In previous sections, we discussed the production of the PBH from the AD baryogenesis assuming that a single inflation with $N_e \gtrsim 60$ is responsible for all scales of our universe.
Although LIGO PBHs are sufficiently produced, so many residual HBBs, which are too small to collapse into the PBHs, are predicted at the same time.
Interestingly they properly contribute to the dark matter abundance in the case of the gauge-mediation.
However, in the case of the gravity-mediation their abundance is constrained not to change the abundances  of the light elements produced by the BBN.
Such excessive residual HBBs arise from the (approximate) scale invariance of the HBB spectrum.
As is discussed in the Sec.~\ref{3B}, the HBBs are produced from the begging of inflation $(N=0)$.
In order to produce the sufficient HBBs with the LIGO scale ($N\sim30$), the coefficient of the Hubble induced mass-term $c_I$ should be much smaller than unity, that is, the HBB spectrum is nearly scale invariant.
This is because for larger $c_I$, the growth of the fluctuation saturates immediately and smaller HBBs are not produced.
Consequently, due to the scale invariance of the spectrum, the residual HBBs are produced as well as the LIGO scale HBBs.

Considering the multi-stage inflation models where inflations occur more than once, we can relax such difficulties.
For example, let us consider the following double inflation scenario with two stages of inflation.
The first inflation with $N_e \ll 60$ produces density perturbations at the CMB scale while the second one responsible for small-scale perturbations.
During the first inflation, we assume that AD field has a large Hubble induced mass ($c_{1I}\gg1$) and its quantum fluctuations do not grow.
On the other hand, the Hubble induced mass of the AD field is assume to be $c_{2I}\lesssim 1$ for the second inflation.
Here $c_{iI}$ is the coefficient of the Hubble induced mass term during the $i$-th inflation.
Then, the HBBs are start to be created at the beginning of the second inflation, which can be much later than the time CMB scale exit the horizon.
After a while, the production of the HBBs saturates due to the Hubble induced mass.
Therefore, the HBBs are produced only around the scale corresponding to the beginning of the second inflation.
If we assume the second inflation starts at $N\sim 30$ when the LIGO PBH scale exits the horizon, we can suppress the excessive residual HBBs.

We can easily apply the calculation in the previous section to the double inflation scenario.
We only have to redefine the parameters as
\begin{align}
   N\equiv \ln(a/a_{i_2}),~~~~c_I\equiv c_{2I},
\end{align}
where $a_{i_2}=a(t_{i_2})$ is the scale factor at the beginning of the second inflation.
The difference is that $N_{\rm CMB}$ takes a negative value because CMB scale exit the horizon during the first inflation.
Here we note that the relation $N_e-N_{\rm CMB}\sim50-60$ should still be satisfied.
We show the PBH abundance in this double inflation scenario in the Fig.~\ref{fig:d}.
Owing to the suppression of the residual HBBs, we can realize the higher peak with the same $f_B$.

\begin{figure}[t]
\begin{center}
  \begin{tabular}{c}
    \begin{minipage}{0.5\hsize}
      \begin{center}
        \includegraphics[clip, width=70mm]{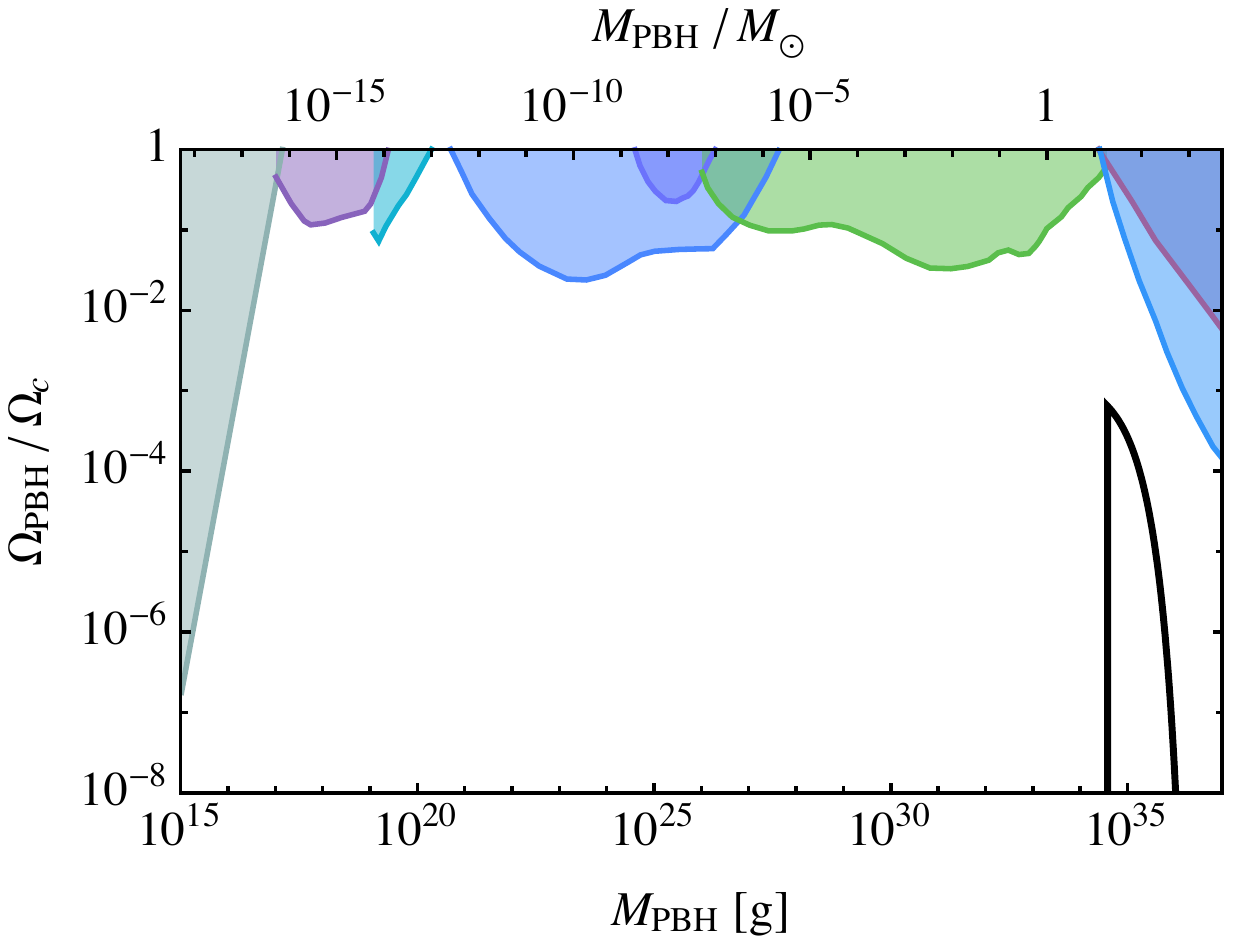}
        \hspace{1.6cm}
              \end{center}
    \end{minipage}
    \begin{minipage}{0.5\hsize}
      \begin{center}
        \includegraphics[clip, width=70mm]{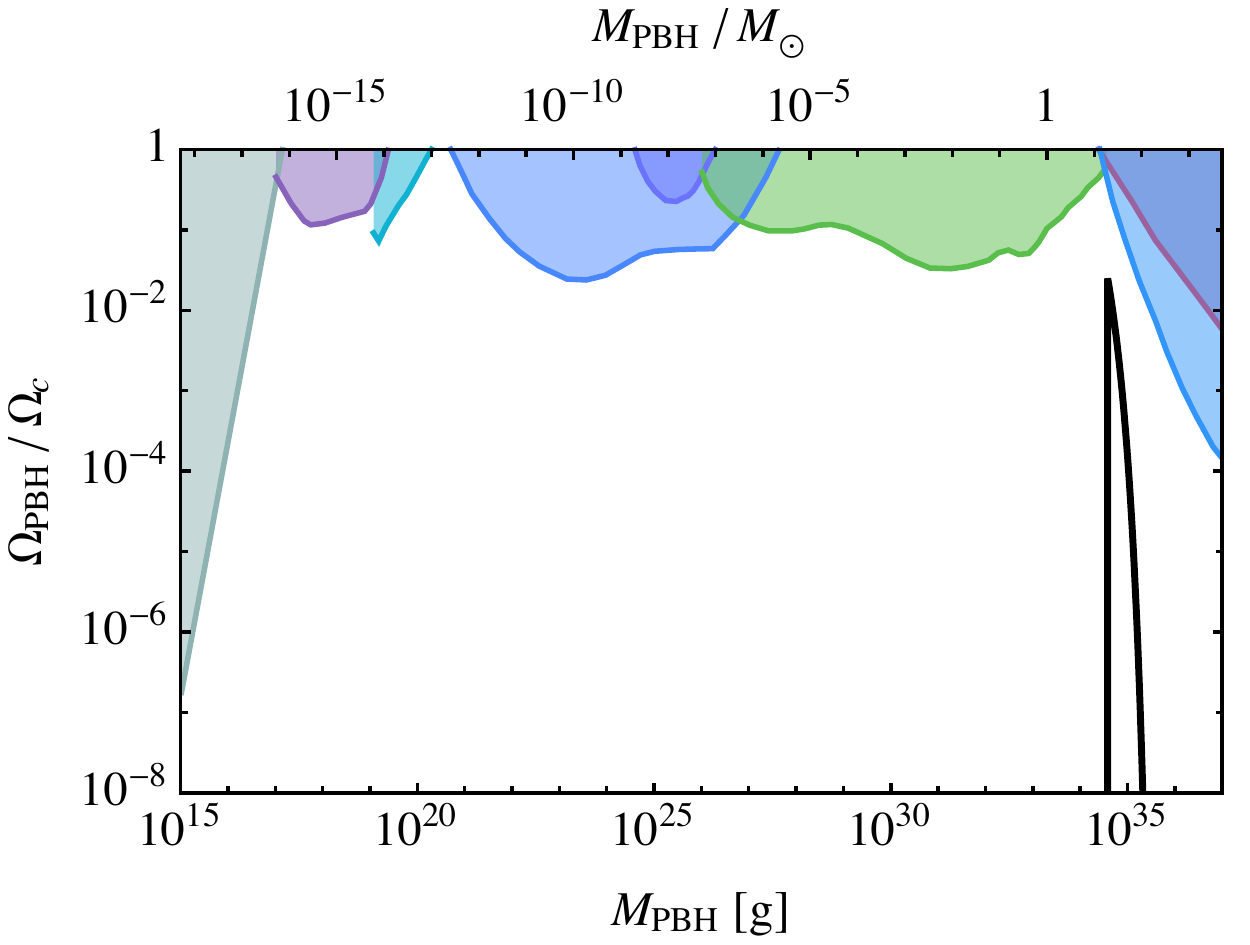}
        \hspace{1.6cm}
      \end{center}
    \end{minipage}

  \end{tabular}
  \caption{We plot the PBH abundance in the double inflation scenario. In the left panel, that in the gravity-mediated SUSY breaking scenario is shown. Here we choose the parameters $(c_I,\Delta,N_{\rm CMB})=(1.22,1.05,-17.5)$ and $\eta_b^{B}/\eta_b^{\rm ob}\simeq10^{-1}$. The left panel is the case in the gauge-mediated SUSY breaking scenario. Here we choose the parameters $(c_I,\Delta,N_{\rm CMB})=(0.92,1.01,-18.5)$ and the dark matter abundance is explained by the residual HBBs. In both model, the peak values get higher than the case in the single inflation scenario.}
  \label{fig:img}
  \end{center}
\end{figure}

Finally, we comment on the supermassive BHs.
In the single inflation scenario, although it successfully produce the LIGO PBHs, one may worry about the over-production of the supermassive BHs.
Due to the scale invariance of the HBB spectrum, the heavier HBBs are abundantly generated as well as HBBs which are responsible for the LIGO PBHs.
As a result, larger amount of the  supermassive BHs are generated (see Fig.~\ref{fig:qcd},\ref{fig:q}).
While they do not conflict with the observational constraints, their abundance is much greater than one in every comoving volume of 1Gpc$^3$ and somewhat unconventional.
On the other hand, in the double inflation scenario, such concern is obviously absent because the HBBs are started to generated only after the second inflation (see Fig.~\ref{fig:img}).

\section{Conclusions and Discussions}
\label{sec:conclusion}
In this paper, we have discussed the formation of the PBH from the AD-mechanism proposed in the ref.\cite{Hasegawa:2017jtk} in more details.
By taking into account that the Hubble induced mass can change before and after inflation, the inhomogeneous AD baryogenesis can take place, which produces HBBs.
The produced HBBs have large density contrasts through the QCD phase transition or Q-ball formation and form PBHs when they reenter the horizon.
This mechanism can explain the LIGO gravitational wave events evading the stringent constraints from the $\mu$-distortion and PTA experiment.
We have considered the gravity- and gauge- mediated SUSY breaking scenarios where the SUSY breaking effect is mediated by the gravity and gauge interactions, respectively.
The SUSY breaking scenarios affect not only the baryon asymmetry inside the HBB but also the properties of the Q-ball, which determines the evolution of the density contrast of the HBBs.

In the case of the gravity-mediated SUSY breaking scenario, the produced Q-balls are unstable against the decay into the baryons, so the baryon number is not confined inside the Q-balls.
Then, the baryon asymmetry in the HBBs is carried by non-relativistic nucleons after the QCD phase transition.
As the universe expands, the density contrast of the HBBs increase and they gravitationally collapse into the PBHs.
The remarkable feature is that the mass spectrum of the PBHs have a lower cut-off because the PBH formation occur only after the QCD phase transition.
Interestingly, this cutoff $M_{\rm QCD}$ coincide with the mass of the LIGO BHs $\sim 30M_{\odot}$. We have shown this mechanism consistently explains the BHs inferred from the LIGO event evading the observational constraints. We note that smaller HBBs which reenter the horizon before the QCD phase transition do not collapse and contribute to the current baryon asymmetry.
Although the formation of the PBH require the huge baryon asymmetry $\eta^{\rm (B)}_b\sim1$, it is naturally realized by the AD mechanism with both $n=4$ and $6$ flat directions.

In the case of the gauge-mediated SUSY breaking scenario, the Q-balls are stable and contribute to the current dark matter.
Thus, the HBBs are eventually dominated by the Q-balls and collapse into the PBHs. The cut-off for the mass spectrum is determined by the horizon size at the Q-ball domination inside the HBBs, which is related to the Q-ball abundance inside the HBBs.
We have shown that if we assume the residual HBBs have a sizable contribution to the dark matter, the sufficient amount of the PBHs are produced so that the event rate of the LIGO events are consistently reproduced.
We call this coincidence as cogenesis of the LIGO PBHs and the dark matter.
Such a large Q-ball abundance is also naturally realized by the AD mechanism with both $n=4$ and $6$ flat directions.

\acknowledgments

We would like to thank Kenta Ando, Jeong-Pyong Hong, Masahiro Ibe, Keisuke Inomata and Eisuke Sonomoto for helpful comments. This work is supported by JSPS KAKENHI Grant Number 17H01131 (M. K.) and 17K05434 (M. K.), MEXT KAKENHI Grant Number 15H05889 (M. K.), JSPS Research Fellowship for Young Scientists Grant Number 17J07391 (F. H.) and also by the World Premier International Research Center Initiative (WPI), MEXT, Japan.


\bibliographystyle{JHEP}
\bibliography{ADPBHf}

\end{document}